\documentclass{article}
\usepackage[utf8]{inputenc}

\usepackage{float,siunitx} 
\usepackage{amsmath,amssymb,amsfonts}
\usepackage{booktabs,caption,subcaption}
\usepackage{algorithmicx,setspace}
\usepackage[noend]{algpseudocode}
\usepackage{graphics,graphicx,xcolor}
\usepackage[utf8]{inputenc}
\usepackage{authblk}
\usepackage{hyperref}
\hypersetup{
    colorlinks=true,
    linkcolor=blue,
    filecolor=magenta,      
    urlcolor=cyan,
    pdftitle={Overleaf Example},
    pdfpagemode=FullScreen,
    }

\newsavebox\affbox

\usepackage{pst-node}
\usepackage{multido}

\usepackage[nameinlink,capitalize]{cleveref}

\usepackage{tikz}
\usetikzlibrary{positioning,chains,fit,shapes,calc}
\usepackage[most]{tcolorbox}
\usepackage{comment}
\usepackage[section]{placeins}

\usepackage{mathtools}

\newcommand{\glabel}[1]{\{X^b\}}

\usepackage{listings}
\newsavebox{\mintedbox}
\usepackage[draft]{minted}
\tcbuselibrary{minted,breakable,skins}

\makeatletter
\def\PYGdefault@reset{\let\PYGdefault@it=\relax \let\PYGdefault@bf=\relax%
    \let\PYGdefault@ul=\relax \let\PYGdefault@tc=\relax%
    \let\PYGdefault@bc=\relax \let\PYGdefault@ff=\relax}
\def\PYGdefault@tok#1{\csname PYGdefault@tok@#1\endcsname}
\def\PYGdefault@toks#1+{\ifx\relax#1\empty\else%
    \PYGdefault@tok{#1}\expandafter\PYGdefault@toks\fi}
\def\PYGdefault@do#1{\PYGdefault@bc{\PYGdefault@tc{\PYGdefault@ul{%
    \PYGdefault@it{\PYGdefault@bf{\PYGdefault@ff{#1}}}}}}}
\def\PYGdefault#1#2{\PYGdefault@reset\PYGdefault@toks#1+\relax+\PYGdefault@do{#2}}

\expandafter\def\csname PYGdefault@tok@w\endcsname{\def\PYGdefault@tc##1{\textcolor[rgb]{0.73,0.73,0.73}{##1}}}
\expandafter\def\csname PYGdefault@tok@c\endcsname{\let\PYGdefault@it=\textit\def\PYGdefault@tc##1{\textcolor[rgb]{0.25,0.50,0.50}{##1}}}
\expandafter\def\csname PYGdefault@tok@cp\endcsname{\def\PYGdefault@tc##1{\textcolor[rgb]{0.74,0.48,0.00}{##1}}}
\expandafter\def\csname PYGdefault@tok@k\endcsname{\let\PYGdefault@bf=\textbf\def\PYGdefault@tc##1{\textcolor[rgb]{0.00,0.50,0.00}{##1}}}
\expandafter\def\csname PYGdefault@tok@kp\endcsname{\def\PYGdefault@tc##1{\textcolor[rgb]{0.00,0.50,0.00}{##1}}}
\expandafter\def\csname PYGdefault@tok@kt\endcsname{\def\PYGdefault@tc##1{\textcolor[rgb]{0.69,0.00,0.25}{##1}}}
\expandafter\def\csname PYGdefault@tok@o\endcsname{\def\PYGdefault@tc##1{\textcolor[rgb]{0.40,0.40,0.40}{##1}}}
\expandafter\def\csname PYGdefault@tok@ow\endcsname{\let\PYGdefault@bf=\textbf\def\PYGdefault@tc##1{\textcolor[rgb]{0.67,0.13,1.00}{##1}}}
\expandafter\def\csname PYGdefault@tok@nb\endcsname{\def\PYGdefault@tc##1{\textcolor[rgb]{0.00,0.50,0.00}{##1}}}
\expandafter\def\csname PYGdefault@tok@nf\endcsname{\def\PYGdefault@tc##1{\textcolor[rgb]{0.00,0.00,1.00}{##1}}}
\expandafter\def\csname PYGdefault@tok@nc\endcsname{\let\PYGdefault@bf=\textbf\def\PYGdefault@tc##1{\textcolor[rgb]{0.00,0.00,1.00}{##1}}}
\expandafter\def\csname PYGdefault@tok@nn\endcsname{\let\PYGdefault@bf=\textbf\def\PYGdefault@tc##1{\textcolor[rgb]{0.00,0.00,1.00}{##1}}}
\expandafter\def\csname PYGdefault@tok@ne\endcsname{\let\PYGdefault@bf=\textbf\def\PYGdefault@tc##1{\textcolor[rgb]{0.82,0.25,0.23}{##1}}}
\expandafter\def\csname PYGdefault@tok@nv\endcsname{\def\PYGdefault@tc##1{\textcolor[rgb]{0.10,0.09,0.49}{##1}}}
\expandafter\def\csname PYGdefault@tok@no\endcsname{\def\PYGdefault@tc##1{\textcolor[rgb]{0.53,0.00,0.00}{##1}}}
\expandafter\def\csname PYGdefault@tok@nl\endcsname{\def\PYGdefault@tc##1{\textcolor[rgb]{0.63,0.63,0.00}{##1}}}
\expandafter\def\csname PYGdefault@tok@ni\endcsname{\let\PYGdefault@bf=\textbf\def\PYGdefault@tc##1{\textcolor[rgb]{0.60,0.60,0.60}{##1}}}
\expandafter\def\csname PYGdefault@tok@na\endcsname{\def\PYGdefault@tc##1{\textcolor[rgb]{0.49,0.56,0.16}{##1}}}
\expandafter\def\csname PYGdefault@tok@nt\endcsname{\let\PYGdefault@bf=\textbf\def\PYGdefault@tc##1{\textcolor[rgb]{0.00,0.50,0.00}{##1}}}
\expandafter\def\csname PYGdefault@tok@nd\endcsname{\def\PYGdefault@tc##1{\textcolor[rgb]{0.67,0.13,1.00}{##1}}}
\expandafter\def\csname PYGdefault@tok@s\endcsname{\def\PYGdefault@tc##1{\textcolor[rgb]{0.73,0.13,0.13}{##1}}}
\expandafter\def\csname PYGdefault@tok@sd\endcsname{\let\PYGdefault@it=\textit\def\PYGdefault@tc##1{\textcolor[rgb]{0.73,0.13,0.13}{##1}}}
\expandafter\def\csname PYGdefault@tok@si\endcsname{\let\PYGdefault@bf=\textbf\def\PYGdefault@tc##1{\textcolor[rgb]{0.73,0.40,0.53}{##1}}}
\expandafter\def\csname PYGdefault@tok@se\endcsname{\let\PYGdefault@bf=\textbf\def\PYGdefault@tc##1{\textcolor[rgb]{0.73,0.40,0.13}{##1}}}
\expandafter\def\csname PYGdefault@tok@sr\endcsname{\def\PYGdefault@tc##1{\textcolor[rgb]{0.73,0.40,0.53}{##1}}}
\expandafter\def\csname PYGdefault@tok@ss\endcsname{\def\PYGdefault@tc##1{\textcolor[rgb]{0.10,0.09,0.49}{##1}}}
\expandafter\def\csname PYGdefault@tok@sx\endcsname{\def\PYGdefault@tc##1{\textcolor[rgb]{0.00,0.50,0.00}{##1}}}
\expandafter\def\csname PYGdefault@tok@m\endcsname{\def\PYGdefault@tc##1{\textcolor[rgb]{0.40,0.40,0.40}{##1}}}
\expandafter\def\csname PYGdefault@tok@gh\endcsname{\let\PYGdefault@bf=\textbf\def\PYGdefault@tc##1{\textcolor[rgb]{0.00,0.00,0.50}{##1}}}
\expandafter\def\csname PYGdefault@tok@gu\endcsname{\let\PYGdefault@bf=\textbf\def\PYGdefault@tc##1{\textcolor[rgb]{0.50,0.00,0.50}{##1}}}
\expandafter\def\csname PYGdefault@tok@gd\endcsname{\def\PYGdefault@tc##1{\textcolor[rgb]{0.63,0.00,0.00}{##1}}}
\expandafter\def\csname PYGdefault@tok@gi\endcsname{\def\PYGdefault@tc##1{\textcolor[rgb]{0.00,0.63,0.00}{##1}}}
\expandafter\def\csname PYGdefault@tok@gr\endcsname{\def\PYGdefault@tc##1{\textcolor[rgb]{1.00,0.00,0.00}{##1}}}
\expandafter\def\csname PYGdefault@tok@ge\endcsname{\let\PYGdefault@it=\textit}
\expandafter\def\csname PYGdefault@tok@gs\endcsname{\let\PYGdefault@bf=\textbf}
\expandafter\def\csname PYGdefault@tok@gp\endcsname{\let\PYGdefault@bf=\textbf\def\PYGdefault@tc##1{\textcolor[rgb]{0.00,0.00,0.50}{##1}}}
\expandafter\def\csname PYGdefault@tok@go\endcsname{\def\PYGdefault@tc##1{\textcolor[rgb]{0.53,0.53,0.53}{##1}}}
\expandafter\def\csname PYGdefault@tok@gt\endcsname{\def\PYGdefault@tc##1{\textcolor[rgb]{0.00,0.27,0.87}{##1}}}
\expandafter\def\csname PYGdefault@tok@err\endcsname{\def\PYGdefault@bc##1{\setlength{\fboxsep}{0pt}\fcolorbox[rgb]{1.00,0.00,0.00}{1,1,1}{\strut ##1}}}
\expandafter\def\csname PYGdefault@tok@kc\endcsname{\let\PYGdefault@bf=\textbf\def\PYGdefault@tc##1{\textcolor[rgb]{0.00,0.50,0.00}{##1}}}
\expandafter\def\csname PYGdefault@tok@kd\endcsname{\let\PYGdefault@bf=\textbf\def\PYGdefault@tc##1{\textcolor[rgb]{0.00,0.50,0.00}{##1}}}
\expandafter\def\csname PYGdefault@tok@kn\endcsname{\let\PYGdefault@bf=\textbf\def\PYGdefault@tc##1{\textcolor[rgb]{0.00,0.50,0.00}{##1}}}
\expandafter\def\csname PYGdefault@tok@kr\endcsname{\let\PYGdefault@bf=\textbf\def\PYGdefault@tc##1{\textcolor[rgb]{0.00,0.50,0.00}{##1}}}
\expandafter\def\csname PYGdefault@tok@bp\endcsname{\def\PYGdefault@tc##1{\textcolor[rgb]{0.00,0.50,0.00}{##1}}}
\expandafter\def\csname PYGdefault@tok@fm\endcsname{\def\PYGdefault@tc##1{\textcolor[rgb]{0.00,0.00,1.00}{##1}}}
\expandafter\def\csname PYGdefault@tok@vc\endcsname{\def\PYGdefault@tc##1{\textcolor[rgb]{0.10,0.09,0.49}{##1}}}
\expandafter\def\csname PYGdefault@tok@vg\endcsname{\def\PYGdefault@tc##1{\textcolor[rgb]{0.10,0.09,0.49}{##1}}}
\expandafter\def\csname PYGdefault@tok@vi\endcsname{\def\PYGdefault@tc##1{\textcolor[rgb]{0.10,0.09,0.49}{##1}}}
\expandafter\def\csname PYGdefault@tok@vm\endcsname{\def\PYGdefault@tc##1{\textcolor[rgb]{0.10,0.09,0.49}{##1}}}
\expandafter\def\csname PYGdefault@tok@sa\endcsname{\def\PYGdefault@tc##1{\textcolor[rgb]{0.73,0.13,0.13}{##1}}}
\expandafter\def\csname PYGdefault@tok@sb\endcsname{\def\PYGdefault@tc##1{\textcolor[rgb]{0.73,0.13,0.13}{##1}}}
\expandafter\def\csname PYGdefault@tok@sc\endcsname{\def\PYGdefault@tc##1{\textcolor[rgb]{0.73,0.13,0.13}{##1}}}
\expandafter\def\csname PYGdefault@tok@dl\endcsname{\def\PYGdefault@tc##1{\textcolor[rgb]{0.73,0.13,0.13}{##1}}}
\expandafter\def\csname PYGdefault@tok@s2\endcsname{\def\PYGdefault@tc##1{\textcolor[rgb]{0.73,0.13,0.13}{##1}}}
\expandafter\def\csname PYGdefault@tok@sh\endcsname{\def\PYGdefault@tc##1{\textcolor[rgb]{0.73,0.13,0.13}{##1}}}
\expandafter\def\csname PYGdefault@tok@s1\endcsname{\def\PYGdefault@tc##1{\textcolor[rgb]{0.73,0.13,0.13}{##1}}}
\expandafter\def\csname PYGdefault@tok@mb\endcsname{\def\PYGdefault@tc##1{\textcolor[rgb]{0.40,0.40,0.40}{##1}}}
\expandafter\def\csname PYGdefault@tok@mf\endcsname{\def\PYGdefault@tc##1{\textcolor[rgb]{0.40,0.40,0.40}{##1}}}
\expandafter\def\csname PYGdefault@tok@mh\endcsname{\def\PYGdefault@tc##1{\textcolor[rgb]{0.40,0.40,0.40}{##1}}}
\expandafter\def\csname PYGdefault@tok@mi\endcsname{\def\PYGdefault@tc##1{\textcolor[rgb]{0.40,0.40,0.40}{##1}}}
\expandafter\def\csname PYGdefault@tok@il\endcsname{\def\PYGdefault@tc##1{\textcolor[rgb]{0.40,0.40,0.40}{##1}}}
\expandafter\def\csname PYGdefault@tok@mo\endcsname{\def\PYGdefault@tc##1{\textcolor[rgb]{0.40,0.40,0.40}{##1}}}
\expandafter\def\csname PYGdefault@tok@ch\endcsname{\let\PYGdefault@it=\textit\def\PYGdefault@tc##1{\textcolor[rgb]{0.25,0.50,0.50}{##1}}}
\expandafter\def\csname PYGdefault@tok@cm\endcsname{\let\PYGdefault@it=\textit\def\PYGdefault@tc##1{\textcolor[rgb]{0.25,0.50,0.50}{##1}}}
\expandafter\def\csname PYGdefault@tok@cpf\endcsname{\let\PYGdefault@it=\textit\def\PYGdefault@tc##1{\textcolor[rgb]{0.25,0.50,0.50}{##1}}}
\expandafter\def\csname PYGdefault@tok@c1\endcsname{\let\PYGdefault@it=\textit\def\PYGdefault@tc##1{\textcolor[rgb]{0.25,0.50,0.50}{##1}}}
\expandafter\def\csname PYGdefault@tok@cs\endcsname{\let\PYGdefault@it=\textit\def\PYGdefault@tc##1{\textcolor[rgb]{0.25,0.50,0.50}{##1}}}

\makeatother

\makeatletter
\def\PYG@reset{\let\PYG@it=\relax \let\PYG@bf=\relax%
    \let\PYG@ul=\relax \let\PYG@tc=\relax%
    \let\PYG@bc=\relax \let\PYG@ff=\relax}
\def\PYG@tok#1{\csname PYG@tok@#1\endcsname}
\def\PYG@toks#1+{\ifx\relax#1\empty\else%
    \PYG@tok{#1}\expandafter\PYG@toks\fi}
\def\PYG@do#1{\PYG@bc{\PYG@tc{\PYG@ul{%
    \PYG@it{\PYG@bf{\PYG@ff{#1}}}}}}}
\def\PYG#1#2{\PYG@reset\PYG@toks#1+\relax+\PYG@do{#2}}

\expandafter\def\csname PYG@tok@w\endcsname{\def\PYG@tc##1{\textcolor[rgb]{0.73,0.73,0.73}{##1}}}
\expandafter\def\csname PYG@tok@c\endcsname{\let\PYG@it=\textit\def\PYG@tc##1{\textcolor[rgb]{0.25,0.50,0.50}{##1}}}
\expandafter\def\csname PYG@tok@cp\endcsname{\def\PYG@tc##1{\textcolor[rgb]{0.74,0.48,0.00}{##1}}}
\expandafter\def\csname PYG@tok@k\endcsname{\let\PYG@bf=\textbf\def\PYG@tc##1{\textcolor[rgb]{0.00,0.50,0.00}{##1}}}
\expandafter\def\csname PYG@tok@kp\endcsname{\def\PYG@tc##1{\textcolor[rgb]{0.00,0.50,0.00}{##1}}}
\expandafter\def\csname PYG@tok@kt\endcsname{\def\PYG@tc##1{\textcolor[rgb]{0.69,0.00,0.25}{##1}}}
\expandafter\def\csname PYG@tok@o\endcsname{\def\PYG@tc##1{\textcolor[rgb]{0.40,0.40,0.40}{##1}}}
\expandafter\def\csname PYG@tok@ow\endcsname{\let\PYG@bf=\textbf\def\PYG@tc##1{\textcolor[rgb]{0.67,0.13,1.00}{##1}}}
\expandafter\def\csname PYG@tok@nb\endcsname{\def\PYG@tc##1{\textcolor[rgb]{0.00,0.50,0.00}{##1}}}
\expandafter\def\csname PYG@tok@nf\endcsname{\def\PYG@tc##1{\textcolor[rgb]{0.00,0.00,1.00}{##1}}}
\expandafter\def\csname PYG@tok@nc\endcsname{\let\PYG@bf=\textbf\def\PYG@tc##1{\textcolor[rgb]{0.00,0.00,1.00}{##1}}}
\expandafter\def\csname PYG@tok@nn\endcsname{\let\PYG@bf=\textbf\def\PYG@tc##1{\textcolor[rgb]{0.00,0.00,1.00}{##1}}}
\expandafter\def\csname PYG@tok@ne\endcsname{\let\PYG@bf=\textbf\def\PYG@tc##1{\textcolor[rgb]{0.82,0.25,0.23}{##1}}}
\expandafter\def\csname PYG@tok@nv\endcsname{\def\PYG@tc##1{\textcolor[rgb]{0.10,0.09,0.49}{##1}}}
\expandafter\def\csname PYG@tok@no\endcsname{\def\PYG@tc##1{\textcolor[rgb]{0.53,0.00,0.00}{##1}}}
\expandafter\def\csname PYG@tok@nl\endcsname{\def\PYG@tc##1{\textcolor[rgb]{0.63,0.63,0.00}{##1}}}
\expandafter\def\csname PYG@tok@ni\endcsname{\let\PYG@bf=\textbf\def\PYG@tc##1{\textcolor[rgb]{0.60,0.60,0.60}{##1}}}
\expandafter\def\csname PYG@tok@na\endcsname{\def\PYG@tc##1{\textcolor[rgb]{0.49,0.56,0.16}{##1}}}
\expandafter\def\csname PYG@tok@nt\endcsname{\let\PYG@bf=\textbf\def\PYG@tc##1{\textcolor[rgb]{0.00,0.50,0.00}{##1}}}
\expandafter\def\csname PYG@tok@nd\endcsname{\def\PYG@tc##1{\textcolor[rgb]{0.67,0.13,1.00}{##1}}}
\expandafter\def\csname PYG@tok@s\endcsname{\def\PYG@tc##1{\textcolor[rgb]{0.73,0.13,0.13}{##1}}}
\expandafter\def\csname PYG@tok@sd\endcsname{\let\PYG@it=\textit\def\PYG@tc##1{\textcolor[rgb]{0.73,0.13,0.13}{##1}}}
\expandafter\def\csname PYG@tok@si\endcsname{\let\PYG@bf=\textbf\def\PYG@tc##1{\textcolor[rgb]{0.73,0.40,0.53}{##1}}}
\expandafter\def\csname PYG@tok@se\endcsname{\let\PYG@bf=\textbf\def\PYG@tc##1{\textcolor[rgb]{0.73,0.40,0.13}{##1}}}
\expandafter\def\csname PYG@tok@sr\endcsname{\def\PYG@tc##1{\textcolor[rgb]{0.73,0.40,0.53}{##1}}}
\expandafter\def\csname PYG@tok@ss\endcsname{\def\PYG@tc##1{\textcolor[rgb]{0.10,0.09,0.49}{##1}}}
\expandafter\def\csname PYG@tok@sx\endcsname{\def\PYG@tc##1{\textcolor[rgb]{0.00,0.50,0.00}{##1}}}
\expandafter\def\csname PYG@tok@m\endcsname{\def\PYG@tc##1{\textcolor[rgb]{0.40,0.40,0.40}{##1}}}
\expandafter\def\csname PYG@tok@gh\endcsname{\let\PYG@bf=\textbf\def\PYG@tc##1{\textcolor[rgb]{0.00,0.00,0.50}{##1}}}
\expandafter\def\csname PYG@tok@gu\endcsname{\let\PYG@bf=\textbf\def\PYG@tc##1{\textcolor[rgb]{0.50,0.00,0.50}{##1}}}
\expandafter\def\csname PYG@tok@gd\endcsname{\def\PYG@tc##1{\textcolor[rgb]{0.63,0.00,0.00}{##1}}}
\expandafter\def\csname PYG@tok@gi\endcsname{\def\PYG@tc##1{\textcolor[rgb]{0.00,0.63,0.00}{##1}}}
\expandafter\def\csname PYG@tok@gr\endcsname{\def\PYG@tc##1{\textcolor[rgb]{1.00,0.00,0.00}{##1}}}
\expandafter\def\csname PYG@tok@ge\endcsname{\let\PYG@it=\textit}
\expandafter\def\csname PYG@tok@gs\endcsname{\let\PYG@bf=\textbf}
\expandafter\def\csname PYG@tok@gp\endcsname{\let\PYG@bf=\textbf\def\PYG@tc##1{\textcolor[rgb]{0.00,0.00,0.50}{##1}}}
\expandafter\def\csname PYG@tok@go\endcsname{\def\PYG@tc##1{\textcolor[rgb]{0.53,0.53,0.53}{##1}}}
\expandafter\def\csname PYG@tok@gt\endcsname{\def\PYG@tc##1{\textcolor[rgb]{0.00,0.27,0.87}{##1}}}
\expandafter\def\csname PYG@tok@err\endcsname{\def\PYG@bc##1{\setlength{\fboxsep}{0pt}\fcolorbox[rgb]{1.00,0.00,0.00}{1,1,1}{\strut ##1}}}
\expandafter\def\csname PYG@tok@kc\endcsname{\let\PYG@bf=\textbf\def\PYG@tc##1{\textcolor[rgb]{0.00,0.50,0.00}{##1}}}
\expandafter\def\csname PYG@tok@kd\endcsname{\let\PYG@bf=\textbf\def\PYG@tc##1{\textcolor[rgb]{0.00,0.50,0.00}{##1}}}
\expandafter\def\csname PYG@tok@kn\endcsname{\let\PYG@bf=\textbf\def\PYG@tc##1{\textcolor[rgb]{0.00,0.50,0.00}{##1}}}
\expandafter\def\csname PYG@tok@kr\endcsname{\let\PYG@bf=\textbf\def\PYG@tc##1{\textcolor[rgb]{0.00,0.50,0.00}{##1}}}
\expandafter\def\csname PYG@tok@bp\endcsname{\def\PYG@tc##1{\textcolor[rgb]{0.00,0.50,0.00}{##1}}}
\expandafter\def\csname PYG@tok@fm\endcsname{\def\PYG@tc##1{\textcolor[rgb]{0.00,0.00,1.00}{##1}}}
\expandafter\def\csname PYG@tok@vc\endcsname{\def\PYG@tc##1{\textcolor[rgb]{0.10,0.09,0.49}{##1}}}
\expandafter\def\csname PYG@tok@vg\endcsname{\def\PYG@tc##1{\textcolor[rgb]{0.10,0.09,0.49}{##1}}}
\expandafter\def\csname PYG@tok@vi\endcsname{\def\PYG@tc##1{\textcolor[rgb]{0.10,0.09,0.49}{##1}}}
\expandafter\def\csname PYG@tok@vm\endcsname{\def\PYG@tc##1{\textcolor[rgb]{0.10,0.09,0.49}{##1}}}
\expandafter\def\csname PYG@tok@sa\endcsname{\def\PYG@tc##1{\textcolor[rgb]{0.73,0.13,0.13}{##1}}}
\expandafter\def\csname PYG@tok@sb\endcsname{\def\PYG@tc##1{\textcolor[rgb]{0.73,0.13,0.13}{##1}}}
\expandafter\def\csname PYG@tok@sc\endcsname{\def\PYG@tc##1{\textcolor[rgb]{0.73,0.13,0.13}{##1}}}
\expandafter\def\csname PYG@tok@dl\endcsname{\def\PYG@tc##1{\textcolor[rgb]{0.73,0.13,0.13}{##1}}}
\expandafter\def\csname PYG@tok@s2\endcsname{\def\PYG@tc##1{\textcolor[rgb]{0.73,0.13,0.13}{##1}}}
\expandafter\def\csname PYG@tok@sh\endcsname{\def\PYG@tc##1{\textcolor[rgb]{0.73,0.13,0.13}{##1}}}
\expandafter\def\csname PYG@tok@s1\endcsname{\def\PYG@tc##1{\textcolor[rgb]{0.73,0.13,0.13}{##1}}}
\expandafter\def\csname PYG@tok@mb\endcsname{\def\PYG@tc##1{\textcolor[rgb]{0.40,0.40,0.40}{##1}}}
\expandafter\def\csname PYG@tok@mf\endcsname{\def\PYG@tc##1{\textcolor[rgb]{0.40,0.40,0.40}{##1}}}
\expandafter\def\csname PYG@tok@mh\endcsname{\def\PYG@tc##1{\textcolor[rgb]{0.40,0.40,0.40}{##1}}}
\expandafter\def\csname PYG@tok@mi\endcsname{\def\PYG@tc##1{\textcolor[rgb]{0.40,0.40,0.40}{##1}}}
\expandafter\def\csname PYG@tok@il\endcsname{\def\PYG@tc##1{\textcolor[rgb]{0.40,0.40,0.40}{##1}}}
\expandafter\def\csname PYG@tok@mo\endcsname{\def\PYG@tc##1{\textcolor[rgb]{0.40,0.40,0.40}{##1}}}
\expandafter\def\csname PYG@tok@ch\endcsname{\let\PYG@it=\textit\def\PYG@tc##1{\textcolor[rgb]{0.25,0.50,0.50}{##1}}}
\expandafter\def\csname PYG@tok@cm\endcsname{\let\PYG@it=\textit\def\PYG@tc##1{\textcolor[rgb]{0.25,0.50,0.50}{##1}}}
\expandafter\def\csname PYG@tok@cpf\endcsname{\let\PYG@it=\textit\def\PYG@tc##1{\textcolor[rgb]{0.25,0.50,0.50}{##1}}}
\expandafter\def\csname PYG@tok@c1\endcsname{\let\PYG@it=\textit\def\PYG@tc##1{\textcolor[rgb]{0.25,0.50,0.50}{##1}}}
\expandafter\def\csname PYG@tok@cs\endcsname{\let\PYG@it=\textit\def\PYG@tc##1{\textcolor[rgb]{0.25,0.50,0.50}{##1}}}

\makeatother

\title{\textbf{ethp2psim}: Evaluating and deploying privacy-enhanced  peer-to-peer routing protocols for the Ethereum network}

\author{Ferenc Béres, István András Seres, Domokos M. Kelen, András A. Benczúr}

\date{}

\begin{document}

\maketitle

\begin{abstract}
Network-level privacy is the Achilles heel of financial privacy in cryptocurrencies. Financial privacy amounts to achieving and maintaining blockchain- and network-level privacy. Blockchain-level privacy recently received substantial attention. Specifically, several privacy-enhancing technologies were proposed and deployed to enhance blockchain-level privacy. On the other hand, network-level privacy, i.e., privacy on the peer-to-peer layer, has seen far less attention and development.

In this work, we aim to provide a peer-to-peer network simulator, \textbf{ethp2psim}, that allows researchers to evaluate the privacy guarantees of privacy-enhanced broadcast and message routing algorithms. Our goal is two-fold. First, we want to enable researchers to implement their proposed protocols in our modular simulator framework. Second, our simulator allows researchers to evaluate the privacy guarantees of privacy-enhanced routing algorithms. Finally, \textbf{ethp2psim} can help choose the right protocol parameters for efficient, robust, and private deployment.
\end{abstract}

\section{Introduction}
Ethereum is the most popular public blockchain according to the number of issued transactions. Ethereum's public ledger is completely transparent: every account's balance and transaction history is visible to everyone. In numerous applications, this level of transparency is undesirable. There are numerous privacy-enhancing technologies described in the literature: mixers~\cite{meiklejohn2018mobius,pertsev2019tornado,seres2019mixeth}, stealth addresses~\cite{todd2014stealth}, and confidential transactions~\cite{bunz2020zether,williamson2018aztec}. Some of these protocols have already been deployed and are continuously serving the Ethereum community by enhancing their privacy on the blockchain~\cite{beres2021blockchain,wang2022zero}. 

However, blockchain privacy alone is not enough to achieve financial privacy. Users should be able to broadcast transactions in a privacy-preserving manner. Specifically, no adversary should be able to link cryptocurrency addresses to their actual users' unique identifiers, e.g., IP addresses, by logging network-level communication between Ethereum full nodes. This problem is even more pronounced in proof-of-stake Ethereum, where consensus participants must regularly broadcast privacy-critical messages such as blocks and attestations. Previous work has already shown various attacks on privacy that exploited peer-to-peer (P2P) information~\cite{biryukov2014deanonymisation,fanti2017anonymity,koshy2014analysis}. To enhance the privacy of cryptocurrency users, several privacy-enhanced broadcast and routing algorithms were proposed~\cite{bojja2017dandelion,fanti2018dandelion++,modinger2018towards}. 

In this work, we provide the following contributions.
\begin{itemize}
    \item We develop an open-source simulator, \textbf{ethp2psim}, that allows anyone to implement and evaluate various privacy-enhanced broadcast and message routing algorithms.
    \item Using our simulator, we evaluate the privacy guarantees of existing privacy-enhanced message routing algorithms on the Ethereum P2P layer.
    \item We identify several trade-offs between privacy, robustness, and efficiency. The quantification of these trade-offs can inform the deployment of these protocols. 
\end{itemize}

The rest of this paper is organized as follows. In Section~\ref{sec:relatedwork}, we present the related work on privacy on the P2P layer of cryptocurrencies. In Section~\ref{sec:background}, we introduce the pertinent background knowledge on privacy-enhanced message routing algorithms. Section~\ref{sec:model} describes our system and threat model of Ethereum's P2P layer. We introduce \textbf{ethp2psim} in Section~\ref{sec:ethp2psim} and evaluate state-of-the-art routing protocols in Section~\ref{sec:evaluation}. 
Finally, we conclude our work in Section~\ref{sec:conclusion}.

\section{Related work}\label{sec:relatedwork}

\subsection{Privacy-enhanced transaction broadcasting}
Dandelion~\cite{bojja2017dandelion} and Dandelion++~\cite{fanti2018dandelion++} are the two most important related works. In Dandelion(++), users (nodes in the P2P graph) do not directly broadcast their transactions. Rather, each user flips a biased coin, and with probability $p$, they broadcast their transaction, while with probability $1-p$, they forward their message to a single randomly selected neighbor. Dandelion(++) offers a privacy-enhanced routing algorithm with minimal implementation complexity. Specifically, Dandelion is implemented and deployed on the Monero network. In our work, we also adapted ideas from the mixnet literature, specifically onion routing~\cite{reed1998anonymous}, and applied it in the context of PoS Ethereum.

\subsection{Evaluating privacy-enhanced routing protocols}
Serena et al. observed that privacy-preserving routing algorithms incur larger delays and less robustness on message delivery~\cite{serena2021simulation}.
Sharma et al. evaluated the anonymity guarantees of several anonymity schemes used by cryptocurrencies (e.g., Dandelion and Dandelion++) and found they provide little to no anonymity guarantees~\cite{sharma2022anonymity}. In their work, they also build a simulator to evaluate previous routing protocols. However, their simulator is not modular and does not consider numerous Ethereum-specific subtleties that severely change the anonymity analysis.

\section{Preliminaries}\label{sec:background}
In this section, we present the pertinent background knowledge on Ethereum's P2P protocol.
\subsection{Participants}
We assume that the following participants run full nodes in the Ethereum P2P network. These participants can be cast into two main categories: validators (i.e., attestors, aggregators, block proposers) and regular users. The main difference is that regular users do not participate in the consensus algorithm. Hence, they only broadcast regular transactions and not consensus-critical messages. For each type of participant, our goal is to devise a protocol that allows no adversary to link the message (e.g., attestation, block, or transaction) to its originator. 
\paragraph{Attestors.} Every validator must broadcast an attestation in each epoch once in a specific slot.
\paragraph{Aggregators.} Designated attestors collect several attestations, i.e., BLS signatures, and batch them. Afterward, they forward the batched signatures to their neighbors
\paragraph{Block proposers.} Validators are selected to propose new blocks according to their stake amount. Since the stake distribution is public information, we cannot hope to achieve a larger degree of anonymity than the Shannon entropy of the stake distribution.
\paragraph{Users.} Users occasionally broadcast transactions. They are not necessarily validators. 

\subsection{The topology of the Ethereum P2P network}
Ethereum's P2P network is a dynamically changing graph. It is a permissionless network, i.e., nodes can join or leave it whenever they want. Therefore, obtaining a precise description of the Ethereum P2P graph at any given time is nearly impossible. Our simulations rely on this dynamically evolving graph's publicly available static measurements (i.e., snapshots).

In its simplest form, one might assume that the Ethereum P2P graph is a random regular graph since the default Ethereum client randomly selects a fixed number of peers to connect with. However, the real graph might look different as users can modify their client software's P2P settings. We model Ethereum's P2P graph as a weighted, directed graph, where weights on the edges denote the pairwise latencies on that particular connection. Gencer et al.~\cite{gencer2018decentralization} measured the peer-to-peer latency of Ethereum peers. They found that Ethereum's P2P latencies follow a distribution with a $171ms$ average and a standard deviation of $76ms$.  The Ethereum P2P graph is measured and described by Cortes and Bautista~\cite{cortes2021discovering}. They found that the Ethereum2 P2P graph is a heavily centralized network that exhibits a spoke-hub distribution. We also obtained a snapshot of the Goerli testnet, which we believe is sufficiently similar to the mainnet P2P topology. In our simulations, we used the aforementioned results of these measurement studies to approximately model Ethereum's P2P network.

\section{Modelling Ethereum's P2P layer}\label{sec:model}

\subsection{Threat Model}
We assume an active local adversary, i.e., an adversary can only observe communication that passes through them. It is well-known that one can only have anonymous communication against active global adversary if mixnets are employed. In the case of Ethereum, the applicability of mixnets is questionable as they incur considerable latencies that are detrimental to achieving consensus. Therefore, we do not consider the global active adversary model. 

\subsection{Anonymity Goals}
We want to achieve the following anonymity goals by applying privacy-enhanced message routing protocols.
\paragraph{Unlinkability}
Our main goal is that the adversary should not be able to link on-chain (e.g., Ethereum addresses) and off-chain (e.g., IP addresses) identities. A stronger anonymity notion would be sender unlinkability, i.e., the adversary should not be able to tell whether two messages were sent by the same user or not. Currently, there is no message routing protocol that would achieve this stronger notion of privacy.
\paragraph{Robustness}
A privacy-enhanced message routing protocol must be robust. In particular, messages must reach the vast majority of the network even if a considerable fraction of the network is adversarial, and timeliness should not be impacted by possible network issues in a small number of nodes.

\paragraph{Latency}
Ethereum's PoS consensus protocol demands low latency of every participant. Participants with high network latency might miss rewards or worse may be punished by not broadcasting certain messages, i.e., inactivity leak.

\paragraph{Long-term feasibility} Any privacy-enhancing technique should be studied by considering its properties on long time scales, as participants might use the same off- and on-chain identities for years.

\paragraph{Spam prevention}
When messages are forwarded in plain text, it is relatively easy to filter out spam messages. However, if we want to move towards routing protocols that forward encrypted messages, then it is more challenging to fend off spam messages. Even in the latter case, we want to prevent spam messages from being forwarded on the P2P layer.

\paragraph{Ease of implementation}
Ideally, we want to apply a message routing protocol that is easy to implement and deploy, as implementation complexity inherently implies an increased attack surface.

\section{P2P privacy solutions}
\subsection{Dandelion and Dandelion++}
Dandelion~\cite{bojja2017dandelion} and Dandelion++~\cite{fanti2018dandelion++} are the primary hop-by-hop routing algorithm to enhance privacy on the P2P level. In a nutshell, in both protocols, messages can be in two phases: stemming (or sometimes anonymity) and spreading (sometimes broadcast) phase. In the stemming phase, nodes flip a biased coin and decide whether to forward the message to a random neighbor, i.e., continue with the stem phase or broadcast the message. Dandelion is implemented and deployed by the Monero cryptocurrency. Recently, Sharma et al.~\cite{sharma2022anonymity} showed that these protocols provide little to no anonymity guarantees in most realistic scenarios.

Even worse, for both of these protocols, there also exists an Ethereum-specific timing side-channels attack. In this attack, an adversary might be able to infer its position in the anonymity phase of a given message in the knowledge of slot starting times.
This valuable information can lead to a loss of anonymity in these protocols. 
\subsection{Onion routing}
Our solution to Ethereum P2P network-level privacy is an onion routing based message protocol that we detail in a separate document~\cite{kelen2023integrated}. Here we remark that a crucial difference between hop-by-hop routing and onion routing is that in onion routing, all messages are encrypted during the anonymity phase, while they are plain-text during the broadcast phase. The advantage of this approach is that the adversary cannot know whether two cyphertexts belong to the same message. However, major challenges are robustness, latency, and spam protection.  

An onion routing inspired P2P privacy protocol for Ethereum can take two paths. First, it would either use the public Tor network~\cite{kaiserd_2022} 
or implement and deploy an integrated Tor-like overlay network on top of the currently existing Ethereum P2P graph. There are pros and cons. We summarize theoretical anonymity guarantees and implementation challenges in more depth in a separate document~\cite{kelen2023integrated}.

\section{\textbf{ethp2psim}: a peer-to-peer network simulator for Ethereum}\label{sec:ethp2psim}
In this section, we describe in detail our \textbf{ethp2psim} simulator. The simulator is open-source and available at \url{https://github.com/ferencberes/ethp2psim}. This simulator aims to enable researchers to implement privacy-enhanced message routing protocols and evaluate their anonymity guarantees in the context of PoS Ethereum.

Our simulator is written in Python and follows object-oriented design patterns. The simulator consists of the following main classes that are also described in the related documentation\footnote{\url{https://ethp2psim.readthedocs.io/en/latest/?badge=latest}}.

\subsection{Network}
This class defines the P2P network in which users can evaluate their protocols. The default P2P graph is regular (see Listing~\ref{code:message}). However, the simulator can work with user-defined P2P graphs as well. Currently, we use a P2P graph obtained from the Goerli testnet as shown in Listing~\ref{code:network}.

\begin{minted}[linenos,fontsize=\small,xleftmargin=0.5cm,numbersep=3pt,frame=lines]{python}
from ethp2psim.network import *
nw_gen = NodeWeightGenerator('stake')
ew_gen = EdgeWeightGenerator('normal')
goerli = GoerliTestnet()
net = Network(nw_gen, ew_gen, graph=goerli.graph)
\end{minted}
\captionof{listing}{Initialize a P2P network based on Goerli testnet data. Channel latencies (edge weights) are sampled from a normal distribution, while node weights are proportional to the distribution of staked Ether values.}\label{code:network}

\subsection{Message}
This class describes the messages that are forwarded and broadcast between P2P participants. We think of messages as attestations, blocks, or transactions. We abstract away the metadata of messages, e.g., message type or size.

\subsection{Protocol}
This class defines and implements the various message-spreading protocols we want to evaluate. Currently, we implemented and considered a simple broadcast to all algorithms, Dandelion~\cite{bojja2017dandelion}, Dandelion++~\cite{fanti2018dandelion++}, and a variant of our Onion routing based algorithm~\cite{kelen2023integrated}.

Since our simulator is highly modular (see Listing~\ref{code:message}), we hope to receive contributions from the community to evaluate novel message-spreading algorithms.

\begin{minted}[linenos,fontsize=\small,xleftmargin=0.5cm,numbersep=3pt,frame=lines]{python}
from ethp2psim.network import *
from ethp2psim.protocols import BroadcastProtocol
from ethp2psim.adversary import Adversary
nw_gen = NodeWeightGenerator('stake')
ew_gen = EdgeWeightGenerator('normal')
# random 3 regular graph with 10 nodes
net = Network(nw_gen, ew_gen, 10, 3)
protocol = BroadcastProtocol(net, broadcast_mode='all')
# adversary controls a random 10% of P2P network nodes
adversary = Adversary(protocol, 0.1)
# message originating from node 0
msg = Message(0)
msg.process(adversary)
# message was sent to 3 neighbors of node 0
print(len(msg.queue))
# output: 3
\end{minted}
\captionof{listing}{First, we initialize a random regular P2P network along with a broadcast to all neighbor protocol and the related adversary. Next, we create a message originating from node $0$, and we start to propagate it on the P2P network by calling the \emph{msg.process} function.}\label{code:message}

\subsection{Adversary}
This class contains the implementation of various adversarial strategies aiming to reduce anonymity guarantees provided by the implemented routing protocols. At the time of writing, we have implemented several general- and protocol-specific adversarial strategies.

Regarding the positioning of the corrupted nodes, the adversary might corrupt nodes randomly in the P2P graph or selectively target central nodes (e.g., nodes with a high degree or Betweenness) as seen in Listing~\ref{code:adversary}. Therefore, we can evaluate which faults cause more significant anonymity losses.

\begin{minted}[linenos,fontsize=\small,xleftmargin=0.5cm,numbersep=3pt,frame=lines]{python}
from ethp2psim.network import *
from ethp2psim.protocols import BroadcastProtocol
from ethp2psim.adversary import Adversary
seed = 42
# Generate a random Barabási-Albert graph with 20 nodes
G = nx.barabasi_albert_graph(20, 3, seed=seed)
nw_gen = NodeWeightGenerator('stake')
ew_gen = EdgeWeightGenerator('normal')
# use the Barabási-Albert graph as a P2P network
net = Network(nw_gen, ew_gen, graph=G, seed=seed)
protocol = BroadcastProtocol(net, broadcast_mode='all', seed=seed)
# select 4 highest degree nodes to be adversaries
adv_nodes = net.get_central_nodes(4, 'degree')
# initialize adversary
adversary = Adversary(protocol, adversaries=adv_nodes, seed=seed)
print(adversary.nodes)
# output: [5, 0, 4, 6]
\end{minted}
\captionof{listing}{Four nodes with the highest degree are set to be adversarial nodes from a random Barabási-Albert graph\protect\footnotemark. Note that by setting the random seed, you can get reproducible results.}\label{code:adversary}

\footnotetext{\url{https://networkx.org/documentation/stable/reference/generated/networkx.generators.random_graphs.barabasi_albert_graph.html}}

\subsection{Simulator \& Evaluator}
The simulator class defines the experiments we wish to run on a specific P2P graph topology with a particular routing algorithm against a specific adversarial strategy. After simulating multiple messages, we can use the Evaluator to calculate the deanonymization power of the adversary with respect to various performance metrics:

\begin{itemize}
    \item \textbf{hit ratio:} The fraction of messages where the adversary correctly identified the originator.
    \item \textbf{inverse rank:} We take the average of $\frac{1}{r_m}$ values over the messages where $r_m$ is the rank of the originator for a given message $m$ in the adversary's ordered list of possible candidates for the originator of $m$.
    \item \textbf{entropy:} the average of Shannon entropies over all messages. 
    \item \textbf{NDCG}: the average normalized discounted cumulative gain~\cite{al2007relationship} over all messages. Basically, we take the average of $\frac{1}{log_{2}(1+r_{m})}$ as only the originator is considered to be the single relevant hit among the candidates.
\end{itemize}

The generated report includes the average ratio (\textit{message\_spread\_ratio}) of nodes reached by messages in the P2P network.

\begin{minted}[linenos,fontsize=\small,xleftmargin=0.5cm,numbersep=3pt,frame=lines]{python}
from ethp2psim.network import *
from ethp2psim.protocols import DandelionProtocol
from ethp2psim.adversary import DandelionAdversary
from ethp2psim.simulator import Simulator, Evaluator
seed = 42
# Sample staked ether amounts as node weights
nw_gen = NodeWeightGenerator('stake')
# Sample channel latencies from a normal distribution
ew_gen = EdgeWeightGenerator('normal')
net = Network(nw_gen, ew_gen, num_nodes=100, k=20, seed=seed)
# A message is broadcasted with a 40% probability in the stem (anonymity) phase of Dandelion
protocol = DandelionProtocol(net, 0.4, broadcast_mode="sqrt", seed=seed)
# Adversary controls 10% of all nodes and it is in the knowledge of the line graph
adversary = DandelionAdversary(protocol, 0.1, active=False, seed=seed)
# 20 random messages are simulated where originators are selected with respect to their stakes
simulator = Simulator(adversary, num_msg=20, use_node_weights=True, verbose=False, seed=seed)
simulator.run()
# The first sent heuristic is used to determine the first broadcaster for each message
evaluator = Evaluator(simulator, estimator='first_sent')
print(evaluator.get_report())
# output: {'estimator': 'first_sent', 'hit_ratio': 0.2, 'inverse_rank': 0.35,
# 'entropy': 2.05, 'ndcg': 0.48, 'message_spread_ratio': 1.0}
\end{minted}
\captionof{listing}{Evaluating the efficacy of an adversary controling 10\% of P2P network nodes against Dandelion. The main goal of the adversary is to determine the originator node for each $20$ simulated message.}\label{code:simulation}

\section{Evaluation}\label{sec:evaluation}

In this section, we evaluate the privacy guarantees of various network anonymity protocols. Furthermore, we analyze the effect of different network topologies and active adversaries on privacy and robustness. We detail the following four experimental measurements.

\subsection{Deanonymization performance}
First, let's start with a simple experiment where we compare the deanonymization power of the adversary when it uses the first-reach or the first-sent heuristics to determine the originator for each message. These estimator strategies are used to guess the first node that broadcasted a given message based on the observations of all adversarial nodes. In short, an adversary using the first-reach heuristic predicts a node to be the first broadcaster if it is the first node that it heard the message from. On the other hand, using the channel latency information of adjacent channels, a first-sent estimator tries to identify the neighbor that first sent the message to any of the adversarial nodes. Naturally, the two predictions might not coincide as the triangle inequality does not necessarily hold for P2P network latency.

\begin{figure}[th!]
    \centering
    \includegraphics[width=\textwidth]{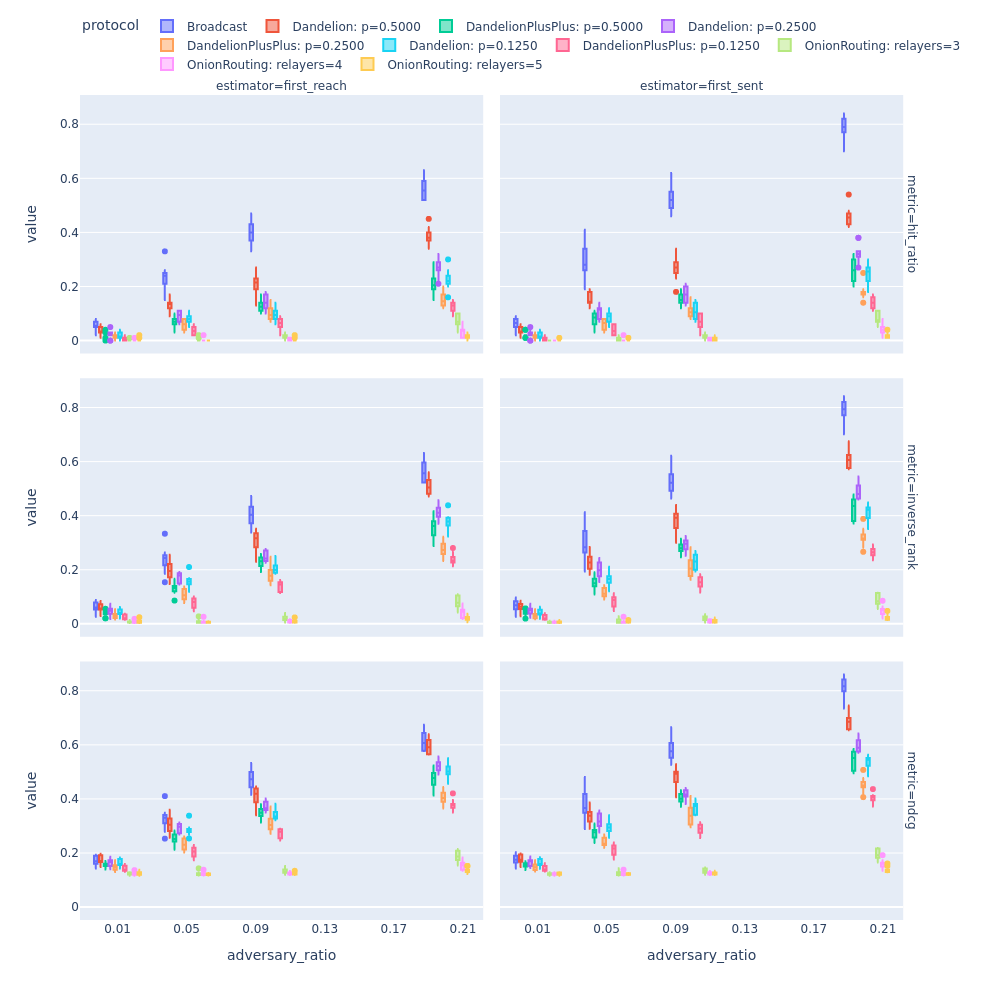}
    \caption{Evaluation of the first sent and first reach estimators for three performance metrics (i.e., hit ratio, inverse rank, NDCG) with respect to multiple protocols (colors) and adversary ratios (x-axis).}
    \label{fig:passive_estimator_check}
\end{figure}

In this experiment, we use a random $50$ regular graph with 1000 nodes to compare the two heuristics against multiple protocols. Not surprisingly, our results in Figure~\ref{fig:passive_estimator_check} show that the adversary with the first-sent estimator performs significantly better. However, we highlight that only the hit ratio, inverse rank, and normalized discounted cumulative gain~\cite{al2007relationship} (NDCG) can reflect this behavior where ground truth information about message sources is compiled into the evaluation.  

\begin{figure}[th!]
    \centering
    \includegraphics[width=\textwidth]{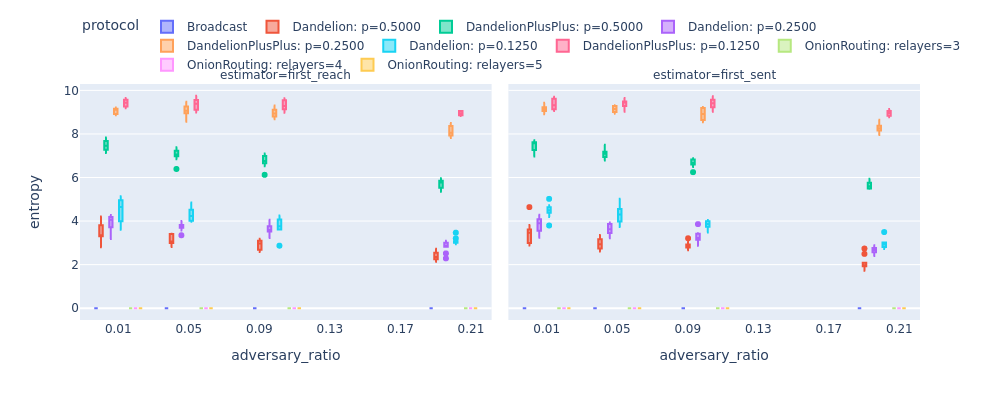}
    \caption{The Shannon entropy of the first sent and first reach estimators for various message spreading algorithms.}
    \label{fig:passive_estimator_entropy}
\end{figure}

Unfortunately, entropy does not depend on the ground truth. It only measures the uncertainty of the predicted distribution but not its closeness to the ground truth. Nevertheless, the entropy for Dandelion++ is higher than for Dandelion in Figure~\ref{fig:passive_estimator_entropy}. The prediction entropy for broadcast to all and our onion routing based protocol is zero as the predicted distribution only contains the most likely candidate. A possible future work could include additional less-likely candidates as well in the prediction distribution, this way better reflecting the knowledge of the adversary.

In Figure~\ref{fig:passive_estimator_check}, it is interesting to see how Dandelion can confuse the adversary compared to simple broadcasting in terms of hit ratio (e.g., first-sent performance drops from $0.5$ to $0.3$ in case of $10\%$ adversarial nodes) which might indicate that it is overly restrictive, as it doesn't contain information about much of the predicted distribution. Instead, \emph{our recommendation is to use inverse rank or NDCG for evaluation}. These metrics can better reflect that despite the higher uncertainty introduced by Dandelion(++), the adversary can still make a good educated guess in the knowledge of the current anonymity graph (i.e., line-graph for Dandelion). For example, in Figure~\ref{fig:passive_estimator_check}, it is quite shocking to see the change in inverse rank from $0.5$ to $0.4$, which means that on average Dandelion improves only half a rank for the predicted message source, in the case of $10\%$ adversarial nodes.

\subsection{Different network topologies}
\begin{figure}[th!]
    \centering
    \includegraphics[width=\textwidth]{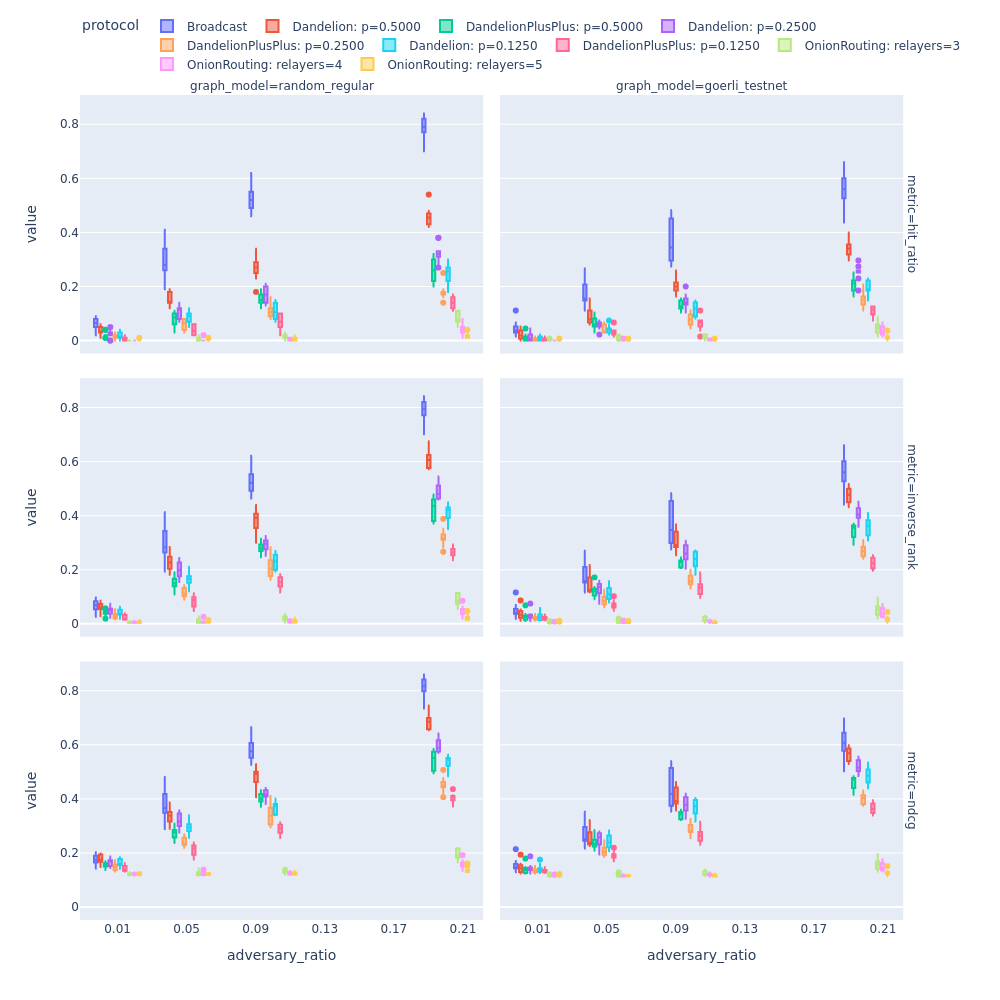}
    \caption{The effect of different network topologies on the adversaries' deanonymization power.}
    \label{fig:different_network_topologies}
\end{figure}

In Figure~\ref{fig:different_network_topologies}, we observe how different graph topologies, such as a random regular graph and a scale-free graph (Goerli testnet's topology), affect the adversary's deanonymization power measured by various different metrics (e.g., hit ratio, inverse rank, NDCG). The deanonymization performance is displayed with respect to the ratio of adversarial nodes (see the x-axis) in the P2P network.

Here, we make four main observations related to privacy:

\begin{itemize}
    \item The achieved privacy is quite brittle in the case of $0.2$ adversary ratio: $0.5$ $\mathit{inverse\_rank}$ for Dandelion with 0.5 broadcast probability means that the adversary outputs a vector of candidates and on average, the true originator is put to the 2nd place.
    \item Dandelion(++) with the least broadcasting probability ($p=0.125$) provides the highest privacy among the considered Dandelion(++)-style protocols.
    \item The results are promising for our onion routing based protocol where the efficiency of the adversary is less affected by the ratio of adversarial nodes in the P2P network.
    \item In general, the Goerli testnet exhibits more privacy across all metrics. 
    
\end{itemize}


\subsection{Broadcast settings}

\begin{figure}[t!]
    \centering
    \includegraphics[width=\textwidth]{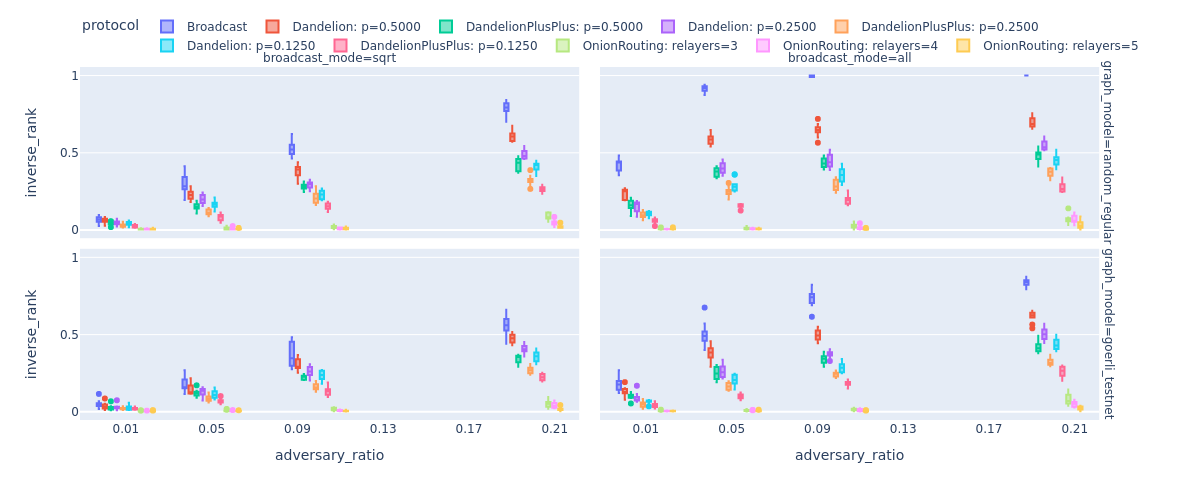}
    \caption{The effect of different broadcast settings on the adversaries' deanonymization power measured in inverse rank.}
    \label{fig:different_broadcast_settings}
\end{figure}

Next, in Figure~\ref{fig:different_broadcast_settings}, observe the significant change in the results when a message is propagated to all neighbors instead of a random square root of them (as we did it in former experiments) during the broadcast phase. It is quite shocking that an adversary controlling $10\%$ of all nodes can be almost sure about the identity of the message source in the case of a simple broadcast protocol. Clearly, Dandelion(++) can significantly decrease the deanonymization performance of the adversary, but it has a high price in terms of robustness, detailed in the next section.

\subsection{Robustness against active and passive adversaries}

\begin{figure}[h!]
    \centering
    \includegraphics[width=\textwidth]{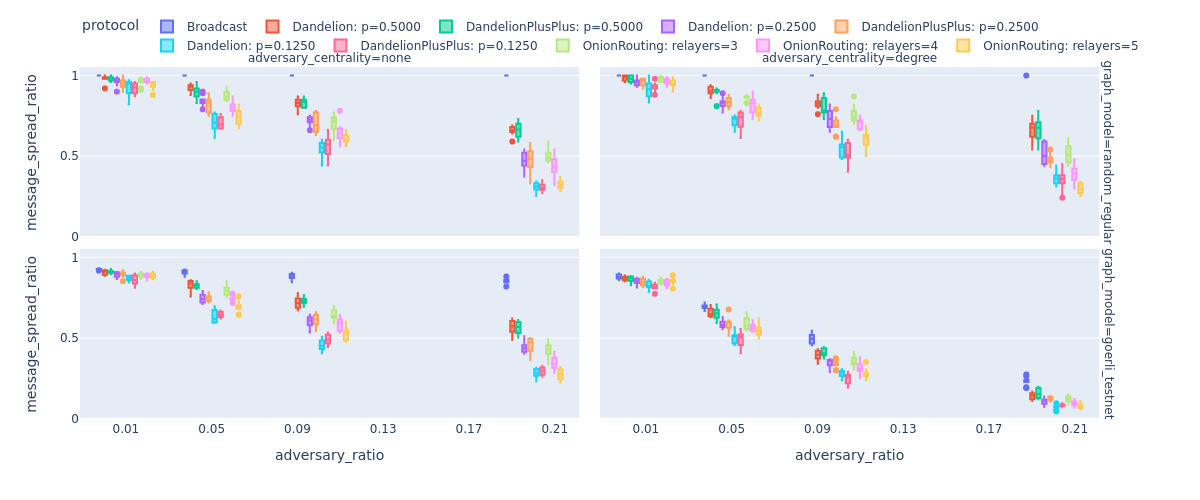}
    \caption{Messages reach fewer nodes if central nodes actively block them.}
    \label{fig:passive_vs_active_adversary}
\end{figure}

In our next experiment, we consider two types of adversaries. A passive adversary that follows the protocol and only logs the timestamp information when its nodes encounter messages. We also implemented an active adversary that does not forward messages at all. In Figure~\ref{fig:passive_vs_active_adversary}, we show that this is especially problematic for Dandelion(++). Imagine that an active adversary sits in the stem (anonymity) phase of Dandelion(++). Basically, if a message encounters an adversarial node on the line graph, then it will never be broadcasted. The more and more adversaries censor messages, the larger the portion of messages that are not heard by nodes in the P2P network. This is even more concerning when the high-degree nodes are compromised (e.g., $\mathit{adversary\_centrality}$=’degree’). Note that the random regular graph is more robust against (active) adversaries than the Goerli testnet.

\begin{figure}[h!]
    \centering    \includegraphics[width=\textwidth]{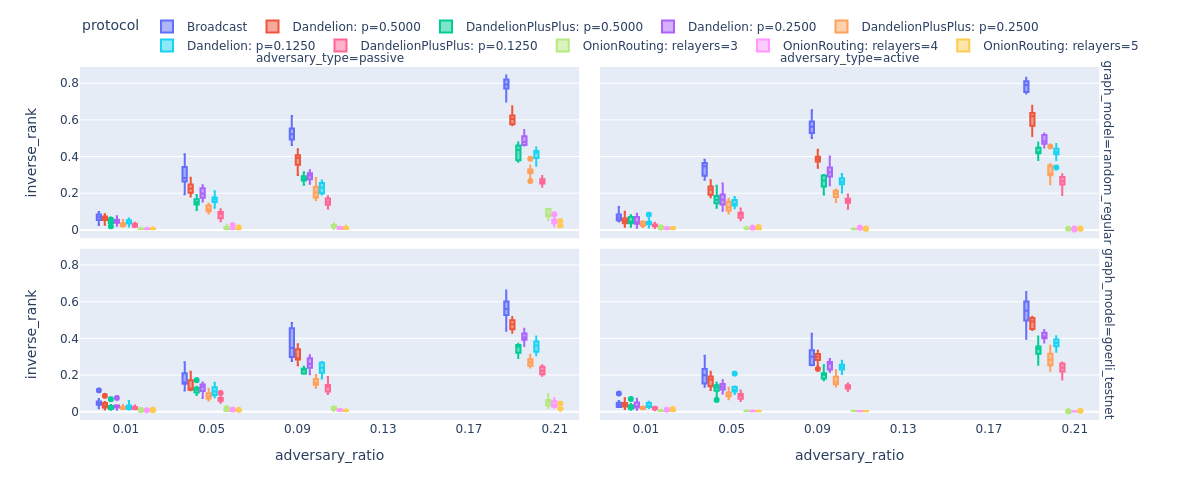}
    \caption{Active and passive adversaries have the same power to deanonymize messages. Deanonymization results are slightly better for the Goerli testnet’s topology.}
    \label{fig:passive_vs_active_performance}
\end{figure}

In Figure~\ref{fig:passive_vs_active_performance}, once again, we see the low levels of privacy (measured in inverse rank in this figure) provided by various privacy-enhanced routing algorithms. It is easy to see that in our setting, active and passive adversaries have the same power to deanonymize messages. Deanonymization results are slightly better for the Goerli testnet’s topology, i.e., the adversary is less powerful on a scale-free graph. In our experiments, the random regular graph has a higher edge density. Hence, the adversary can make a more informed guess about the originator of the messages.

\section{Conclusion}\label{sec:conclusion}
In this work, we developed \textbf{ethp2psim} a network privacy simulator of Ethereum's P2P layer that allows the development, testing, and evaluation of privacy-enhanced message routing algorithms such as Dandelion(++), onion routing, or any custom-specified algorithm. We considered several existing privacy-preserving routing algorithms and evaluated their privacy guarantees using this simulator.

Our simulator is open source and open to contributions. It can be enhanced in several ways.
\begin{itemize}
    \item Due to the modular nature of \textbf{ethp2psim}, one might add new privacy-preserving algorithms and novel adversarial strategies to our package.
    \item We considered several notions of privacy such as Shannon entropy, the ratio of deanonymized messages (\textit{hit\_ratio}), inverse rank, and NDCG. One might also add other measures to evaluate the privacy and anonymity guarantees of P2P message routing algorithms.
    \item Several parameters and protocol internals of Ethereum's proof-of-stake consensus algorithm is simplified. One can refine the simulation by incorporating a more fine-grained implementation of Proof-of-Stake Ethereum's protocol internals, such as network latencies and graph topologies. 
\end{itemize}

We invite the Ethereum community to propose, implement, and evaluate the robustness and privacy guarantees of novel privacy-preserving routing algorithms using \textbf{ethp2psim}.

\paragraph{Acknowledgements}
The development of this simulator and our research were funded by the Ethereum Foundation's Academic Grant Rounds 2022.


\end{document}